\documentclass[twoside]{dis09}
\usepackage[latin1]{inputenc}
\usepackage[dvips]{graphicx,epsfig,color}
\usepackage{wrapfig,rotating}
\usepackage{amssymb,amsmath,array}

\newcommand{\gsim}{\raisebox{-0.07cm  }
{$\, \stackrel{>}{{\scriptstyle\sim}}\, $}}
\newcommand{\ep}{\varepsilon}
\newcommand{\N}{\nonumber}
\newcommand{\D}{\displaystyle}

\newcommand{\MS}{\overline{{\sf MS}}}
\newcommand{\MOM}{\tiny{\mbox{MOM}}}
\begin{document}

\title{\vspace*{-2mm} 
{\tiny DESY 09-103,SFB/CPP-09-66 \hfill IFIC/09-31}\\ 
Moments of the $3$--loop corrections to the heavy flavor contribution to 
$F_2(x,Q^2)$ for $Q^2\gg m^2$.}

\author{
Isabella Bierenbaum$^{1,2}$, Johannes Bl{\"umlein}$^1$ and Sebastian 
Klein$^1$
%
\thanks{This work was supported in part by DFG Sonderforschungsbereich Transregio 9, Computergest\"utzte Theoretische Teilchenphysik, Studienstiftung des Deutschen Volkes, the European Commission MRTN HEPTOOLS under Contract No. MRTN-CT-2006-035505, the Ministerio de Ciencia e Innovacion under Grant No. FPA2007-60323, CPAN (Grant No. CSD2007-00042), the Generalitat Valenciana under Grant No. PROMETEO/2008/069, and by the European Commission MRTN FLAVIAnet under Contract No. MRTN-CT-2006-035482.}
%
\vspace{.3cm}\\
%
1-Deutsches Elektronen--Synchrotron, DESY,\\
Platanenallee 6, D--15738 Zeuthen, Germany
%
\vspace{.1cm}\\
2-Instituto de F\'{i}sica Corpuscular, CSIC-Universitat de Val\`{e}ncia, \\
Apartado de Correos 22085, E-46071 Valencia, Spain.\\
}

\maketitle

\begin{abstract}
\noindent
 We calculate moments of the $O(\alpha_s^3)$ heavy flavor contributions to the 
 Wilson coefficients of the structure function $F_2(x,Q^2)$ in the region
 $Q^2\gg m^2$. The massive Wilson coefficients are obtained as convolutions
 of massive operator matrix elements (OMEs) and the known light flavor Wilson
 coefficients.
 The calculation of moments of the massive OMEs
 involves a first independent recalculation of moments of the
 fermionic contributions to all $3$--loop anomalous dimensions
 of the unpolarized twist--$2$ local composite operators stemming from the
 light--cone expansion \cite{url}.
\end{abstract}

%

\section{Introduction}
\noindent
Deep-inelastic scattering (DIS)
of charged or neutral leptons off proton
and deuteron targets, in the region of large enough values of the gauge boson
virtuality $Q^2 = -q^2$, allows to measure
the leading twist parton densities of the nucleon, the QCD-scale
$\Lambda_{\rm QCD}$, and the strong coupling constant $a_s(Q^2) =
\alpha_s(Q^2)/(4\pi)$, to high precision.
For unpolarized DIS via single photon exchange,
the double--differential cross section
can be expressed in terms of two inclusive
structure functions $F_{2,L}(x,Q^2)$.
These decompose for twist $\tau=2$ into a Mellin convolution of
{\sf non-perturbative} massless parton densities
$f_j(x,\mu^2)$ and the {\sf perturbative}
Wilson coefficients ${\cal C}_{j,(2,L)}(x,Q^2/\mu^2,m_k^2/\mu^2)$.
The latter describe the hard scattering of the photon with a massless 
parton.
They are given by 
the sum of the purely light -- denoted by $C_{j,(2,L)}$ --
and heavy flavor contributions, ${\sf H}_{j,(2,L)}$.
Here $k=c,~b$ and $j=q,~g$, depending on the type of process one considers. 
$x$ denotes the Bjorken scaling variable. 
Especially in the region of smaller
values of Bjorken--$x$, the structure functions contain large
$c\overline{c}$--contributions of up to 20-40~\%, denoted by
$F_{2,L}^{c\overline{c}}(x,Q^2)$.
The perturbative heavy flavor Wilson coefficients
corresponding to these structure functions are known at ${\sf NLO}$
semi--analytically in $x$--space \cite{HEAV1}.
Due to the size of
the heavy flavor corrections, it is necessary to extend the description of
these contributions to $O(a_s^3)$, and thus to the
same level which has been reached for the massless Wilson coefficients \cite{Vermaseren:2005qc}.

A calculation of
these quantities in the whole kinematic range at ${\sf NNLO}$ seems to be out
of reach at present. However, in the limit of large virtualities $Q^2$, $Q^2\:
\gsim \:10\:m_c^2$ in the case of $F_2^{c\bar{c}}(x,Q^2)$, one observes that
$F_{2,L}^{c\bar{c}}(x,Q^2)$ are very well described by their asymptotic
expressions \cite{Buza:1995ie} neglecting power corrections in $m^2/Q^2$.  In this
kinematic range, one can calculate the heavy flavor Wilson coefficients
analytically. This has been done for $F^{c\bar{c}}_2(x,Q^2)$ to 2--loop order
in \cite{Buza:1995ie,BBK1} and for $F^{c\bar{c}}_L(x,Q^2)$ to 3--loop order in
\cite{Blumlein:2006mh}. Note that in the latter case, 
the asymptotic result becomes valid only at
much higher values of $Q^2$. The asymptotic expressions are obtained by 
a factorization of the heavy quark Wilson coefficients into a Mellin 
convolution of massive
OMEs $A_{jk}$ and the massless Wilson coefficients $C_{j,i}$,
if one heavy quark flavor of mass $m$ and $n_f$ light flavors are
considered. 
In the present paper, we report on the calculation of the massive OMEs
$A_{jk}$ to 3--loop order for fixed even moments of the 
Mellin variable $N$, cf. \cite{Bierenbaum:2009mv} for details.
We further calculate the OMEs which are 
required to define heavy quark parton densities in the 
variable flavor number scheme~\cite{Buza:1996wv}.
We also obtain moments of the terms
$\propto T_F$ of the 3--loop unpolarized anomalous dimensions $\gamma_{ij}$.
Our results agree with those obtained in \cite{ANDIMNSS}.
Since the present calculation is completely independent by method, formalism, 
and codes, it provides a strong check on the previous results.
%
%
%
\section{Heavy Flavor Operator Matrix Elements}
\noindent
The heavy flavor Wilson coefficients for a single massive quark may be expressed as
 \begin{eqnarray}
   {\sf H}^{\sf S,PS,NS}_{j,(2,L)}
             \left(x,\frac{Q^2}{\mu^2},\frac{m^2}{\mu^2}\right)
      = 
        H_{j,(2,L)}^{\sf S,PS}
               \left(x,\frac{Q^2}{\mu^2},\frac{m^2}{\mu^2}\right)
      + L_{j,(2,L)}^{\sf S,PS,NS}
             \left(x,\frac{Q^2}{\mu^2},\frac{m^2}{\mu^2}\right)~,
       \label{Callsplit}
  \end{eqnarray}
where the photon couples to a
light $(L)$ or heavy $(H)$ quark line, respectively. 
Further ${\sf S}$ stands for the
flavor--singlet contributions, which are separated into a pure-singlet ${\sf 
(PS)}$ and 
non--singlet (${\sf NS}$) part via ${\sf S}={\sf PS}+{\sf NS}$.
The factorization formula for
the inclusive Wilson coefficients reads in Mellin space, \cite{Buza:1995ie,Buza:1996wv},
 \begin{eqnarray}
     {\cal C}^{{\sf S,PS,NS}, \small{{\sf \sf as}}}_{j,(2,L)}
          \Bigl(N,n_f,\frac{Q^2}{\mu^2},\frac{m^2}{\mu^2}\Bigr) 
        &=& 
           \sum_{i} A^{\sf S,PS,NS}_{ij}\Bigl(N,n_f,\frac{m^2}{\mu^2}\Bigr)
                    C^{\sf S,PS,NS}_{i,(2,L)}
                      \Bigl(N,n_f+1,\frac{Q^2}{\mu^2}\Bigr)
          ~. \label{CallFAC}
 \end{eqnarray}
Here $\mu$ refers to the factorization scale
between the heavy and light contributions in ${\cal {C}}_{j,(2,L)}$ and 
{\sf 'as'} denotes the limit $Q^2\gg~m^2$.
The $C_{j,(2,L)}$ are precisely the light Wilson coefficients and describe 
all the process dependence. The arguments $(n_f),~(n_f+1),$ indicate at how 
many light flavors the respective quantities have to be taken. 
This factorization 
is only valid if the heavy quark coefficient functions are defined in such
a way that all radiative corrections containing heavy quark loops 
are included. Otherwise (\ref{CallFAC}) would not show the correct 
asymptotic $Q^2$--behavior \cite{Buza:1996wv}. 
The mass dependence is given by the process independent massive
OMEs $A_{ij}$, which are the 
flavor--decomposed twist--2 operator matrix elements 
\begin{eqnarray}
  A_{ki}^{\sf S,NS}\Bigl(\frac{m^2}{\mu^2},N\Bigr)  = \langle
  i|O_k^{\sf S, NS}|i\rangle_H
  = \delta_{ki} + \sum_{l=1}^{\infty} a_s^l
     A_{ki}^{{\sf S, NS},(l)}\Bigl(\frac{m^2}{\mu^2},N\Bigr)~.
       \label{OMEs}
\end{eqnarray}
Here, $i$ denotes the external on--shell particle ($i=q,g$) and $O_k$
stands for the quarkonic ($k=q$) or gluonic ($k=g$) operator emerging in the
light--cone expansion. The subscript $H$ indicates that we require the
presence of heavy quarks of one type with mass $m$. 
The logarithmic terms in $m^2/\mu^2$ are completely
determined by renormalization and contain contributions of the anomalous
dimensions of the twist--2 operators.  Thus at ${\sf NNLO}$ the fermionic
parts of the 3-loop anomalous dimensions calculated in Refs. \cite{ANDIMNSS}
appear. All pole terms of the unrenormalized results provide a check on our 
calculation and the single pole terms allow for a first independent calculation
of the terms $\propto T_F$ of the 3-loop anomalous dimensions. 

In case of the gluon operator, the contributing terms are denoted by
$A_{gq,Q}$ and $A_{gg,Q}$. For the quark operator, one distinguishes whether 
the operator couples to a heavy or light quark. In the ${\sf NS}$--case, the 
operator, by definition, couples to the light quark. Thus there is only one 
term, $A_{qq,Q}^{\sf NS}$. In the ${\sf S}$ and ${\sf PS}$--case, two OMEs can 
be distinguished, $\D{\{A_{qq,Q}^{\sf PS},~A_{qg,Q}^{\sf S}\}}$
and $\D{\{A_{Qq}^{\sf PS},~A_{Qg}^{\sf S}\}}$, where, in the former case, 
the operator couples to a light quark and in the latter case to a heavy 
quark. 

Eq. (\ref{CallFAC}) allows to calculate the heavy flavor Wilson
coefficients
in the limit $Q^2\gg m^2$ up to $O(a_s^3)$ by combining the results obtained
in Ref. \cite{Vermaseren:2005qc} for the light flavor Wilson coefficients with the 
$3$--loop massive OMEs which are computed
in this work \cite{Bierenbaum:2009mv}.

A related application of the heavy OMEs is given when using a variable flavor
number scheme to describe parton densities including massive quarks. The OMEs
are then the transition functions going from $n_f$ to $n_f+1$ flavors. One
thus may define parton densities for massive quarks, see
e.g. Ref. \cite{Buza:1996wv}. This is of particular interest for heavy quark
induced processes at the LHC, such as $c\overline{s}\rightarrow W^+$ at large
enough scales $Q^2$.  
%
%
%
\section{Renormalization}
\noindent
We work in Feynman gauge and use dimensional regularization in $D=4+\ep$
dimensions, applying the $\overline{\rm{MS}}$--scheme, if not stated
otherwise. The renormalization proceeds in four steps, which we will briefly
sketch here and refer to \cite{Bierenbaum:2009mv} for more details. 
Mass renormalization is performed in
the on--shell scheme \cite{MASS2}, whereas for charge renormalization we use
the $\overline{\rm{MS}}$--scheme. We 
work in an intermediate ${\sf MOM}$--scheme for charge renormalization by 
requiring that the heavy quark loop contributions to the gluon propagator 
vanish for on--shell external momentum. This is necessary for the 
renormalization 
of the massive OMEs to cancel infrared singularities which would otherwise 
remain. $Z_g$ in this ${\sf MOM}$--scheme can be calculated using the 
background field method \cite{Abbott:1980hw}. 
Finally, we transform our result back to the $\MS$--scheme 
for coupling constant renormalization via, \cite{Bierenbaum:2009mv},
 \begin{eqnarray}
   a_s^{\MOM}&=& a_s^{\MS}
                -\beta_{0,Q}\ln \Bigl(\frac{m^2}{\mu^2}\Bigr) {a_s^{\MS}}^2
                +\Biggl[ \beta^2_{0,Q}\ln^2 \Bigl(\frac{m^2}{\mu^2}\Bigr) 
                        -\beta_{1,Q}\ln \Bigl(\frac{m^2}{\mu^2}\Bigr) 
                        -\beta_{1,Q}^{(1)}
                 \Biggr] {a_s^{\MS}}^3~, \label{asmoma}
  \end{eqnarray}
with 
  \begin{eqnarray}
   \beta_{0,Q} &=&-\frac{4}{3}T_F~, \quad
   \beta_{1,Q} =
                  - 4 \left(\frac{5}{3} C_A + C_F \right) T_F~, \quad
   \beta_{1,Q}^{(1)}=
                           -\frac{32}{9}T_FC_A
                           +15T_FC_F~.
 \label{b1Q1}
  \end{eqnarray}
The remaining singularities are of the
ultraviolet and collinear type. The former are renormalized via the operator
$Z$--factors, whereas the latter are removed via mass factorization through
the transition functions $\Gamma$.
After coupling-- and mass renormalization, the renormalized heavy flavor OMEs
are then obtained by
\begin{equation}
 A=Z^{-1} \hat{A}  \Gamma^{-1}~, \label{GenRen}
\end{equation}
where quantities with a hat are unrenormalized. Note that in the singlet case
operator mixing occurs and hence Eq. (\ref{GenRen}) should be read as a matrix
equation, contrary to the ${\sf NS}$--case.  The $Z$-- and $\Gamma$--factors
can be expressed in terms of the anomalous dimensions of the twist--$2$
operators to all orders in the strong coupling constant,
cf. \cite{Bierenbaum:2009mv,Bierenbaum:2008yu} up to $O(a_s^3)$. 
From Eq. (\ref{GenRen}) one can infer that for operator
renormalization and mass factorization at $O(a_s^3)$, the anomalous dimensions
up to ${\sf NNLO}$ \cite{ANDIMNSS} together with the $1$--loop massive
OMEs up to $O(\ep^2)$ and the $2$--loop massive OMEs up to $O(\ep)$ are
needed. The $2$--loop OMEs up to $O(\ep^0)$ were calculated in Refs. 
\cite{Buza:1995ie,BBK1,Buza:1996wv,Bierenbaum:2009zt}. Higher orders in $\ep$ enter 
since they multiply $Z-$ and
$\Gamma$--factors containing poles in $\ep$. This has been worked out in 
detail in Ref. \cite{Bierenbaum:2008yu}, where we presented the $O(\ep)$ terms
$\overline{a}_{Qg}^{(2)}$, $\overline{a}_{qq,Q}^{(2), {\sf NS}}$ and
$\overline{a}_{Qq}^{(2){\sf PS}}$. The terms
$\overline{a}_{gg,Q}^{(2)}$ and $\overline{a}_{gq,Q}^{(2)}$ were given in
Refs. \cite{Bierenbaum:2009zt}. Thus all terms needed for the renormalization 
at $3$--loops in the unpolarized case are known.

\begin{figure}[htb]
      \begin{center}
       \includegraphics[angle=0, width=2.5cm]{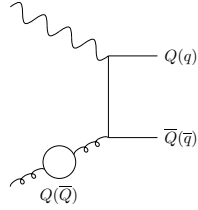}
      \end{center}
       \caption{\sf $O(a_s^2)$ virtual heavy quark corrections to ${\sf H}_{g,(2,L)}^{(2)}$.}
       \label{2LOOPIRR}
     \end{figure}

Finally we would like to point out the difference between the 
${\sf MOM}$-- and $\MS$--scheme for coupling constant renormalization
at ${\sf NLO}$ \cite{Bierenbaum:2009zt}.
Eq. (\ref{CallFAC}) holds only
for completely inclusive quantities, including radiative corrections 
containing heavy quark loops \cite{Buza:1996wv}.
Additionally, (\ref{CallFAC})
has to be applied in such a way that renormalization of the coupling constant 
is carried out in the same scheme for all quantities contributing, i.e., the 
$\MS$--scheme. If one evaluates the 
heavy-quark Wilson coefficients, diagrams
of the type shown in Fig. \ref{2LOOPIRR} may appear as well. 
It contains a virtual heavy quark loop correction to
the gluon propagator in the initial state and contributes to the terms 
$L_{g,i}$ and $H_{g,i}$, respectively, depending on whether a light or 
heavy quark pair is produced in the final state.
Note that in the former case, this diagram contributes to $F_{(2,L)}(x,Q^2)$ in
the inclusive case, but is absent in the semi--inclusive 
$Q\overline{Q}$--production cross section.
In Refs.~\cite{HEAV1}, the coupling
constant was renormalized in the ${\sf MOM}$--scheme in $O(a_s^2)$ by 
absorbing 
the contributions of the above diagram into the coupling constant.
This can be made explicit by considering the 
complete gluonic Wilson coefficient up to $O(a_s^2)$, including one heavy 
quark, see Eq. (\ref{CallFAC}),
\begin{eqnarray}
      &&\hspace{-5mm}C_{g,2}(n_f)+L_{g,2}(n_f+1)+H_{g,2}(n_f+1) 
      =
           a_s^{\MS} \Bigl[~A_{Qg}^{(1), \MS}~
                     +C^{(1)}_{g,2}(n_f+1) \Bigr]
\N\\ &&\hspace{-5mm}
         + {a_s^{\MS}}^2 \Bigl[~A_{Qg}^{(2), \MS}~
     +A_{Qg}^{(1), \MS}~C^{(1), {\sf NS}}_{q,2}(n_f+1)
     +A_{gg,Q}^{(1), \MS}~C^{(1)}_{g,2}(n_f+1) 
     +C^{(2)}_{g,2}(n_f+1) \Bigr]~.
  \label{Sec5Ex1}
\end{eqnarray}
The above equation is given in the $\overline{\sf MS}$--scheme.
Here, the diagram shown in Fig. \ref{2LOOPIRR} contributes, 
corresponding exactly 
to the color factor $T_F^2$. Transformation to the
${\sf MOM}$--scheme for $a_s$, Eq.~(\ref{asmoma}), yields 
\begin{eqnarray}
      &&C_{g,2}(n_f)+L_{g,2}(n_f+1)+H_{g,2}(n_f+1) 
      =
           a_s^{\MOM} \Bigl[~A_{Qg}^{(1), \MOM}~
                     +C^{(1)}_{g,2}(n_f+1) \Bigr]
\N\\ &&\hspace{2mm}
         + {a_s^{\MOM}}^2 \Bigl[~A_{Qg}^{(2), \MOM}~
     +A_{Qg}^{(1), \MOM}~C^{(1), {\sf NS}}_{q,2}(n_f+1)
     +C^{(2)}_{g,2}(n_f+1) \Bigr]~.  \label{Sec5Ex3}
\end{eqnarray}
In the above equation, all contributions due to diagram \ref{2LOOPIRR} have 
canceled, i.e. the color factor $T_F^2$ does not occur
at the $2$--loop level in the ${\sf MOM}$--scheme.
Splitting up Eq.~(\ref{Sec5Ex3}) into $H_{g,i}$ and $L_{g,i}$, 
one observes that $L_{g,i}$  vanishes at $O(a_s^2)$. The term 
$H_{g,i}$ is the one calculated in Ref.~\cite{Buza:1995ie}, which is 
the asymptotic expression of the gluonic heavy flavor Wilson coefficient 
as calculated exactly in Refs.~\cite{HEAV1}. 
It is not clear whether the same can be achieved at the $3$--loop level as 
well, i.e., transforming the general inclusive factorization 
formula (\ref{CallFAC}) in such a way that only the contributions due to 
heavy flavors in the final state remain. Therefore one should use the
asymptotic expressions at $3$ loops only for completely inclusive analyzes.
This approach has also been adopted in 
Ref.~\cite{Buza:1996wv} for the renormalization of the massive OMEs, which 
was performed in the $\overline{\sf MS}$--scheme
and not in the ${\sf MOM}$--scheme, as previously in Ref.~\cite{Buza:1995ie}.
In the ${\sf NS}$-case a similar argument holds, which can be found in
Ref.~\cite{Buza:1995ie}.
%
%
%
\section{Calculation and Results}
\noindent
The massive OMEs at $O(a_s^3)$ are given by $3$--loop self--energy type
diagrams, which contain a local operator insertion. The external massless
particles are on--shell. The heavy quark mass sets the scale and the spin of
the local operator is given by the Mellin--variable $N$. The steps for the
calculation are the following: We use {\sf QGRAF} \cite{Nogueira:1991ex} for the
generation of diagrams.  Approximately $2700$ diagrams contribute to all 
the OMEs.
For the calculation of the color factors we refer 
to \cite{vanRitbergen:1998pn}.
The diagrams are then genuinely given as tensor integrals. 
Applying a suitable projector provides the results for
the specific Mellin moment under consideration. The diagrams are
further translated into a form, which is suitable for the program ${\sf
MATAD}$ \cite{Steinhauser:2000ry}, through which the expansion in $\ep$ is performed and
the corresponding massive three--loop tadpole--type diagrams are
calculated. We have implemented all these steps into a {\sf FORM}--program
\cite{Vermaseren:2000nd} and checked our procedures against various complete two--loop
results and certain scalar $3$--loop integrals and found full agreement.

Applying Eq. (\ref{GenRen}), one can predict the pole structure of the
unrenormalized results and thus the logarithmic terms of the renormalized
OMEs. These contributions can be expressed in terms of the anomalous 
dimensions 
up to $3$ loops, the expansion coefficients of the QCD 
$\beta$--function up to $2$ loops and the $1$-- and $2$--loop 
contributions to the massive OMEs. Thus 
the logarithmic terms are known for general values of $N$. This is not 
the case for the constant term, which contains the genuine $3$--loop 
contributions $a_{ij}^{(3)}$. These are known for the fixed values of $N$
as calculated in this work \cite{Bierenbaum:2009mv}.

For the OMEs $A_{Qg}^{(3)}, A_{qg,Q}^{(3)}$ and 
$A_{gg,Q}^{(3)}$ the moments $N = 2$ to 10, for $A_{Qq}^{(3), \rm PS}$ to 
$N = 12$, and for 
$A_{qq,Q}^{(3), \rm NS}$, $A_{qq,Q}^{(3),\rm PS}$, $A_{gq,Q}^{(3)}$ to $N=14$ 
were computed.
For the flavor non-singlet terms, we calculated as well the odd moments $N=1$ 
to $13$, corresponding to the light flavor \mbox{$-$-combinations}. 
The complete 
calculation took about $250$ days of computer time.
All our results agree with the predictions obtained from renormalization, 
providing us with a strong check on our calculation. As an example, we show 
the constant term of ${A_{gq,Q}}$ for $N=2$
\begin{eqnarray}
   A_{gq,Q}^{(3), \MS}\Bigg|_{\mu^2=m^2}^{N=2} &=&
T_FC_F \Biggl\{
C_A
      \Bigl( 
                 \frac{101114}{2187}
                -\frac{128}{9}{\sf B_4}+128\zeta_4
                -\frac{8296}{81}\zeta_3
      \Bigr)
+T_F
      \Bigl( 
                -\frac{26056}{729}
\N\\ && \hspace{-35mm}
                +\frac{1792}{27}\zeta_3
      \Bigr)
+n_fT_F
      \Bigl( 
                 \frac{44272}{729}
                -\frac{1024}{27}\zeta_3
      \Bigr)
+C_F
      \Bigl( 
                -\frac{570878}{2187}
                +\frac{256}{9}{\sf B_4}-128\zeta_4
                +\frac{17168}{81}\zeta_3
      \Bigr)
    \Biggr\}
~. 
 \label{AgqQ32} 
\end{eqnarray}
Here $\zeta_i$ denotes the Riemann $\zeta$--function at integer argument 
$i$ and 
the term $B_4$ is given by
 \begin{eqnarray}
  {\sf B_4}&=&-4\zeta_2\ln^2 2 +\frac{2}{3}\ln^4 2 -\frac{13}{2}\zeta_4
             +16 {\sf Li}_4\Bigl(\frac{1}{2}\Bigr)~.
 \label{B4}
  \end{eqnarray}
It appears in all OMEs we calculated and is known to arise as 
a genuine mass effect.
%
%
%
%
\section{Conclusions and Outlook}
\noindent
We calculated all massive $3$--loop OMEs for even Mellin--moments
$N=2...10(12,14)$ using {\sf MATAD}. This confirms for the
first time, in an independent calculation, the moments of the fermionic parts 
of
the
corresponding $3$--loop anomalous dimensions \cite{ANDIMNSS}. Combining 
our results with \cite{Vermaseren:2005qc}, this provides fixed moments of the 
heavy flavor Wilson coefficients of $F_2$
in the limit $Q^2\gg m^2$.
First phenomenological studies of the effects of our calculation are 
in preparation.
\section*{Acknowledgments}
We would like to thank K. Chetyrkin, J. Smith, M. Steinhauser and 
J. Vermaseren for  useful discussions. We thank 
both IT groups of DESY providing us access to special facilities to perform 
the present calculation.
\begin{footnotesize}

\end{footnotesize}
\end{document}